\documentclass[acmsmall,screen]{acmart}
\AtBeginDocument{%
  \providecommand\BibTeX{{%
    \normalfont B\kern-0.5em{\scshape i\kern-0.25em b}\kern-0.8em\TeX}}}

\usepackage{float}

\usepackage{booktabs}
\usepackage{array}
\usepackage{balance} 
\usepackage{multirow}
\usepackage[normalem]{ulem}
\usepackage{color}
\definecolor{lightgray}{RGB}{215,215,215}
\usepackage{colortbl}  
\usepackage{xcolor}
\useunder{\uline}{\ul}{}
\usepackage{subfigure}
\usepackage{algorithm}  
\usepackage{algorithmicx}  
\usepackage[noend]{algpseudocode}  

\usepackage{amsmath}  
\usepackage{enumitem}
\usepackage{tabularx}
\usepackage[utf8]{inputenc}
\usepackage[english]{babel}
\usepackage{amsthm}
\usepackage{bm}
\newcommand{\ie}{\emph{i.e., }}
\newcommand{\eg}{\emph{e.g., }}

\newcommand{\wrt}{\emph{w.r.t. }}

\newlength\myindent
\floatname{algorithm}{Algorithm}

\begin{document}


\title{Generative Recommendation: Towards Next-generation Recommender Paradigm}


\author{Wenjie Wang}
\email{wenjiewang96@gmail.com}
\affiliation{%
  \institution{National University of Singapore}
  \country{Singapore}
}
\author{Xinyu Lin}
\email{xylin1028@gmail.com}
\affiliation{%
  \institution{National University of Singapore}
  \country{Singapore}
}
\author{Fuli Feng}
\authornote{Corresponding author: Fuli Feng (fulifeng93@gmail.com).}
\email{fulifeng93@gmail.com}
\affiliation{%
  \institution{University of Science and Technology of China}
  \country{China}
}
\author{Xiangnan He}
\email{xiangnanhe@gmail.com}
\affiliation{%
  \institution{University of Science and Technology of China}
  \country{China}
}
\author{Tat-Seng Chua}
\email{dcscts@nus.edu.sg}
\affiliation{%
  \institution{National University of Singapore}
  \country{Singapore}
}

\begin{abstract}
Recommender systems typically retrieve items from an item corpus for personalized recommendations. However, such a retrieval-based recommender paradigm faces two limitations: 1) the human-generated items in the corpus might fail to {satisfy} the users' diverse information needs, and 2) users usually adjust the recommendations via passive and inefficient feedback such as clicks. Nowadays, AI-Generated Content (AIGC) has revealed significant success across various domains, offering the potential to overcome these limitations: 1) generative AI can produce personalized items to {satisfy} users' specific information needs, and 2) the newly emerged large language models with strong language understanding and generation abilities significantly reduce the efforts of users to precisely express information needs via natural language instructions. In this light, the boom of AIGC points the way towards the next-generation recommender paradigm with two new objectives: 1) generating personalized content through generative AI, and 2) integrating user instructions to guide content generation.

To this end, we propose a novel \textbf{Gene}rative \textbf{Rec}ommender paradigm named GeneRec, which adopts an AI generator to personalize content generation and leverages user instructions to acquire users' information needs. Specifically, we pre-process users' instructions and traditional feedback (\eg clicks) via an instructor to output the generation guidance. Given the guidance, we instantiate the AI generator through an AI editor and an AI creator to repurpose existing items and create new items, respectively. Eventually, GeneRec can perform content retrieval, repurposing, and creation to {satisfy} users' information needs. Besides, to ensure the trustworthiness of the generated items, we emphasize various fidelity checks such as authenticity and legality checks. {Moreover, we provide a roadmap to envision future developments of GeneRec and we present several domain-specific applications of GeneRec with some potential research tasks.} Lastly, we study the feasibility of implementing the AI editor and AI creator on micro-video generation, showing promising results.
\end{abstract}


\maketitle

\section{Introduction}
\label{sec:introduction}
\begin{figure}[t]
\setlength{\abovecaptionskip}{0.1cm}
\setlength{\belowcaptionskip}{-0.20cm}
\centering
\includegraphics[scale=0.6]{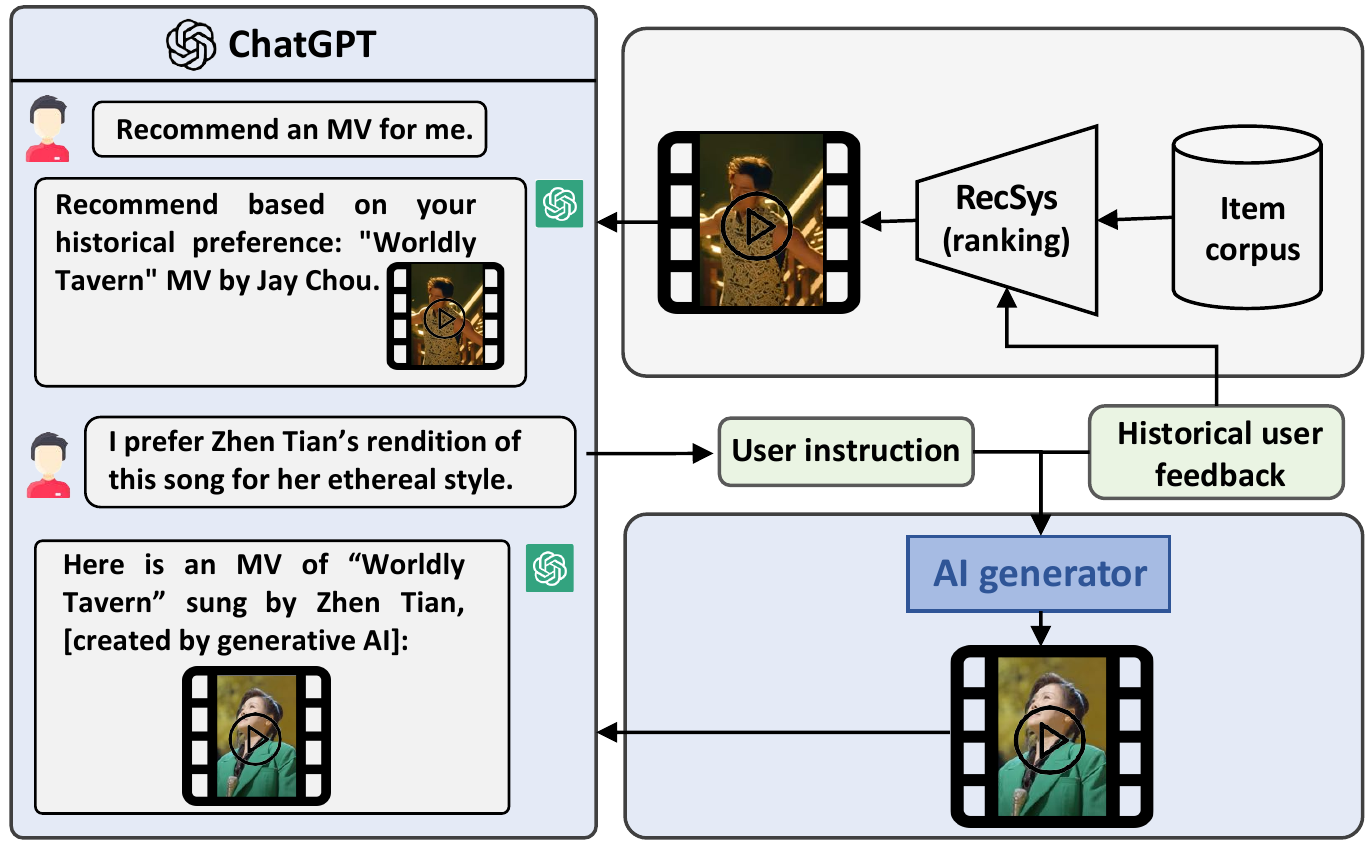}
\caption{An example of using generative AI to interact with users and generate new items in the micro-video domain.}
\label{fig:AIGC_example}
\end{figure}

Recommender systems {fulfill} users' information needs by retrieving item content in a personalized manner. Traditional recommender systems primarily retrieve human-generated content such as {expert-generated movies on Netflix and user-generated micro-videos on Tiktok{~\cite{gomez2015netflix}}.} However, AI-Generated Content (AIGC) has emerged as a prevailing trend across various domains. The advent of powerful neural networks, {exemplified by diffusion models~\cite{rombach2022high}}, has enabled generative AI to produce superhuman content. 
{As shown in Figure~\ref{fig:AIGC_example}, ChatGPT~\cite{ouyang2022training, brown2020language} demonstrates a remarkable ability to interact with users via natural language and generative AI is possible to create multimodal content such as video, text, and audio to form new items.} 
Driven by the boom of AIGC, recommender systems must move beyond human-generated content, {by envisioning a generative recommender paradigm to automatically repurpose existing items\footnote{Here, repurposing means editing existing items for a different purpose, \ie satisfying another user's personalized preference.} or create new items.}


\footnotetext[1]{\url{https://chat.openai.com/chat/}.}
\footnotetext[2]{\url{https://stablediffusionweb.com/}.}

To envision the generative recommender paradigm, we first retrospect the traditional retrieval-based recommender paradigm. The traditional paradigm ranks human-generated items in the item corpus, recommends the top-ranked items to users, and then collects user feedback (\eg clicks) and context (\eg interaction time) to optimize the future rankings for users{~\cite{davidson2010youtube}}. 
Despite its success, such a traditional paradigm suffers from two limitations. 1) The content available in the item corpus might be insufficient to {satisfy} users' personalized information needs. 
{For instance, users may prefer a music video performed by a singer in a specific style (see Figure~\ref{fig:AIGC_example}), while generating such a multimodal music video by humans is impossible or costly{~\cite{wu2023ai}}.} 
And 2) users are currently able to refine the recommendations mostly via passive feedback (\eg clicks), which cannot express their information needs explicitly and efficiently{~\cite{liu2010unifying,liang2023promoting}}. 

AIGC offers the potential to overcome the inherent limitations of the retrieval-based recommender paradigm. 
In particular, 1) {generative AI can generate personalized content to supplement existing items, including repurposing existing items and creating new items{~\cite{brooks2023instructpix2pix,singer2022make}}.} 
Additionally, 2) the newly emerged Large Language Models (LLMs) show strong language understanding and generation abilities{~\cite{wei2022emergent}}, which can effectively reduce users' efforts to convey their diverse information needs via natural language instructions (Figure~\ref{fig:AIGC_example}). 
{Compared to traditional conversational agents, users will engage more readily with advanced ChatGPT-like models, supplementing traditional user feedback.} 
In this light, the emerging AIGC has spurred new objectives for the next-generation recommender systems to enable: 
1) the automatic generation of personalized content through generative AI, and 2) the integration of user instructions to guide content generation. 


To this end, we propose a novel \textbf{Gene}rative \textbf{Rec}ommender paradigm called \textbf{GeneRec}, which integrates the powerful generative AI for personalized content generation, including both repurposing and creation. 
Figure~\ref{fig:paradigm_abs} illustrates how GeneRec adds a loop between \textit{an AI generator} and \textit{users}. 
Taking user instructions and feedback as inputs, the AI generator needs to understand users' information needs and generate personalized content. The generated content can be either added to the item corpus for ranking or directly recommended to the users. 
Wherein, the user instructions are not only limited to textual conversations but can also include multimodal conversations, \ie fusing images, videos, audio, and natural languages to express the information needs. 

To instantiate the GeneRec paradigm, 
{we formulate one instructor module to process the instructions, as well as two modules for item repurposing and creation.
Specifically, the instructor module pre-processes user instructions and feedback to determine whether to initiate the content generation, and also encodes the instructions and feedback to guide the content generation.}
Given the guidance, an AI editor repurposes an existing item to fulfill users' specific preference, \ie personalized item editing, and an AI creator directly creates new items for personalized item creation. 
To ensure the trustworthiness and high quality of generated items, we emphasize the importance of various fidelity checks from the aspects of bias, privacy, safety, authenticity, and legality{~\cite{wu2023ai,guo2023aigc,wang2023security}}. 
{Moreover, in Section~\ref{sec:paradigm}, we present a roadmap to explain the future development trends of GeneRec. Besides, we introduce several application scenarios of GeneRec across different domains in Section~\ref{sec:domain_cases} and detail some potential research tasks under GeneRec in Section~\ref{sec:tasks}.} 
Lastly, to explore the feasibility of applying the 
recent advance in AIGC to implement the AI editor and AI creator, we devise several tasks of micro-video generation and conduct experiments on a high-quality micro-video dataset. Empirical results show that existing AIGC methods can accomplish some repurposing and creation tasks, and it is promising to achieve the grand objectives of GeneRec in the future. We release our code and dataset at \url{https://github.com/Linxyhaha/GeneRec}.



To summarize, our contributions are threefold.
\begin{itemize}[leftmargin=*]
    \item We highlight the essential role of AIGC in recommender systems and point out the extended objectives for next-generation recommender systems: moving towards a generative recommender paradigm, which can naturally interact with users via multimodal instructions, and flexibly retrieve, repurpose, and/or create item content to {meet} users' diverse information needs in various recommendation domains. 
    
    \item {We propose to instantiate the generative recommender paradigm by formulating three key modules:
    the instructor for interacting with users and processing user instructions to guide content generation, the AI editor for personalized item editing, and the AI creator for personalized item creation.}
    
    \item {We spotlight the essential perspectives of fidelity checks and present a roadmap with several application scenarios and potential research tasks to envision the future directions of GeneRec.}
    
    \item {We investigate the feasibility of utilizing existing AIGC methods to implement the AI editor and AI creator in the micro-video recommendation domain.}
    
    
\end{itemize}

\section{Generative Recommender Paradigm}
\label{sec:paradigm}

\begin{figure}[t]
\setlength{\abovecaptionskip}{0.2cm}
\setlength{\belowcaptionskip}{-0.20cm}
\centering
\includegraphics[scale=0.42]{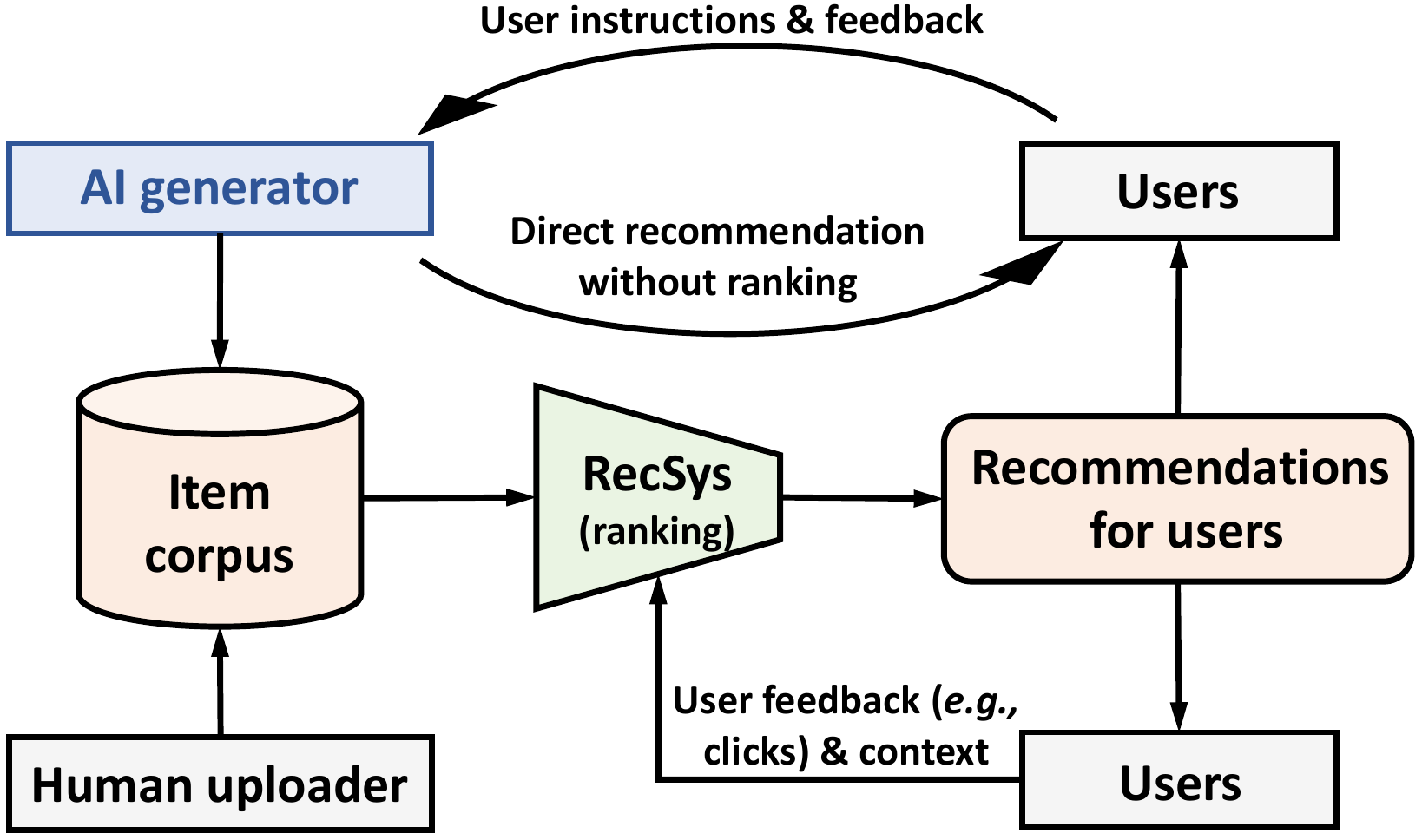}
\caption{Illustration of the GeneRec paradigm. The AI generator takes user instructions and feedback to generate personalized content, which can be directly recommended to users or fed to the item corpus for ranking with human-generated items.}
\label{fig:paradigm_abs}
\end{figure}

{We propose two new objectives for the next-generation recommender systems: 1) automatically repurposing or creating items via generative AI, and 2) integrating rich user instructions.} 
To achieve these objectives, we present GeneRec to complement the traditional retrieval-based recommender paradigm. 

\vspace{8pt}
\noindent$\bullet$ \textbf{Overview.}
Figure~\ref{fig:paradigm_abs} presents the overview of the proposed GeneRec paradigm with two loops. 
In the traditional retrieval-based user-system loop, 
{human uploaders}, including domain experts (\eg musicians) and regular users (\eg micro-video users), generate and upload items to the item corpus. These items are then ranked for recommendations according to the user preference, where the preference is learned from the context (\eg interaction time) and user feedback over historical recommendations.

To complement this traditional paradigm, GeneRec adds another loop between the AI generator and users. Users can control the content generated by the AI generator through user instructions and feedback. 
Thereafter, the generated items can be directly exposed to the users without ranking if the users clearly express their expectations for AI-generated items or if they have rejected human-generated items via negative feedback (\eg dislikes) many times. In addition, the AI-generated items can be ranked together with the human-generated items to output the recommendations.

\vspace{8pt}
\noindent$\bullet$ \textbf{User instructions.}
The strong conversational ability of ChatGPT-like LLMs can enrich the interaction modes between users and the AI generator. The users can flexibly control content generation via conversational instructions, where the instructions can be either textual conversations or multimodal conversations. Through instructions, users can 1) freely enable the AI generator to generate their preferred content at any time, and 2) express their information needs more quickly and efficiently than 
interaction-based feedback. In particular, users cannot only express what they like, but can also indicate what they dislike via instructions. In the past, users were unwilling to make efforts to give explicit feedback such as instructions, while LLMs will be incredibly intelligent to collect user instruction by audio or multimodal inputs, reducing the burden of users via advanced interfaces in the future. 

\vspace{8pt}
\noindent$\bullet$ {\textbf{Content generation.}}
Before content generation, the AI generator might need to \textbf{\textit{pre-process}} the user instructions, for instance, some pre-trained language models might require designing prompts~\cite{brown2020language} or instruction tuning~\cite{wei2021finetuned}; diffusion models may need to simplify queries or extract instruction embeddings as inputs for image synthesis~\cite{rombach2022high}. In addition to user instructions, user feedback such as clicks can also guide the content generation since user instructions might ignore some {implicit user preference} and the AI generator can infer such preference from users' historical interactions. {Learning implicit user preference from noisy and passive user feedback (\eg clicks and dwell time) has long been a central focus in the field of recommendations. GeneRec can borrow the prior experience from the traditional recommendation domain for user preference modeling.}

Subsequently, the AI generator learns personalized information needs from user instructions and feedback, and then generates personalized item content accordingly. 
The generation includes both \textbf{\textit{repurposing}} existing items and \textbf{\textit{creating}} new items from scratch. For example, to repurpose a micro-video, we may convert it into multiple styles or split it into clips with different themes for distinct users; besides, the AI generator may select a topic and create a new micro-video based on user instructions and collected Web data (\eg facts and knowledge).

\textbf{\textit{Post-processing}} is essential to ensure the quality of generated content. The AI generator can judge whether the generated content will satisfy users' information needs and further refine it, such as adding captions and subtitles for micro-videos. {The relevance between the generated content and users' information needs is also significant. Besides, it is also vital to ensure the trustworthiness of generated content through fidelity checks. }


\vspace{8pt}
\noindent$\bullet$ \textbf{Fidelity checks.}
To ensure the generated content is accurate, fair, and safe, GeneRec should pass the fidelity checks. {Although different recommendation scenarios may require various fidelity checks, they generally should include but not be limited to the following perspectives.}
\begin{itemize}[leftmargin=*]
    \item [1)] \textbf{Bias and fairness}: the AI generator might learn from biased data~\cite{baeza2020bias}, and thus should confirm that the generated content does not perpetuate stereotypes, promote hate speech and discrimination, cause unfairness to certain populations, or reinforce other harmful biases~\cite{gao2020counteracting, fu2020fairness, zehlike2022fair}. {For example, the news generation, especially regarding sensitive topics, should pay more attention to the checks of bias and fairness to avoid ethical and social issues.}
    
    \item [2)] \textbf{Privacy}: the generated content cannot disseminate any sensitive or personal information that may violate someone's privacy~\cite{shin2018privacy}. {GeneRec may generate an item based on a user's personalized data, and then the generated item may be disseminated to other users, possibly leading to privacy leakage. Many recommendation scenarios such as news, tweets, and micro-videos are sensitive to such privacy leakage.}

    \item [3)] \textbf{Safety}: the AI generator must not pose any risks of harm to users, including the risks of physical and psychological harm~\cite{amodei2016concrete}. For instance, the generated micro-video for teenagers should not contain any unhealthy content. Besides, it is crucial to prevent GeneRec from various attacks such as shilling attack~\cite{gunes2014shilling,chirita2005preventing}. 

    \item [4)] \textbf{Authenticity}: to prevent misinformation spread, we need to verify that the facts, statistics, and claims made in the generated content are accurate based on reliable sources~\cite{maras2019determining}. {Users need to access factual information in certain recommendation scenarios, such as daily events and historical facts, making the authenticity of the generated content extremely important.}

    \item [5)] \textbf{Legal compliance}: more importantly, AIGC must comply with all relevant laws and regulations~\cite{buiten2019towards}. For instance, if the generated micro-videos are about recommending healthy food, they must follow the regulations of healthcare. 
    {Besides, copyright regulation should also be considered to tackle the intricacies of authorship and ownership regarding the content edited or created by the generative AI~\cite{lucchi2023chatgpt}. The government should work with the recommender platform to publish new regulations to clarify the ownership regarding AI-generated content.}

    \item [6)] \textbf{Identifiability}: to assist with AIGC supervision, we suggest adding \textit{digital watermark}~\cite{van1994digital} into AI-generated content for distinguishing human-generated and AI-generated items. {For example, in the fashion domain, designers' patents are essential. Therefore, it is necessary to distinguish between AI-generated, human-generated, and AI-assisted human-generated content.} We can develop AI technologies to automatically identify AI-generated items~\cite{mitchell2023detectgpt}. Furthermore, we may consider deleting the AI-generated items after browsing by users to prevent them from being reused for inappropriate context. 

\end{itemize}

\noindent$\bullet$ \textbf{Evaluation.}
To evaluate the generated content, we propose two kinds of evaluation setups: 1) item-side evaluation and 2) user-side evaluation. \textbf{\textit{Item-side evaluation}} emphasizes the measurements from the item itself, including the item quality measurements (\eg using Fréchet Video Distance (FVD) metric~\cite{voleti2022MCVD} to measure micro-video quality), {the relevance between the generated content and users' information needs, and various fidelity checks}.  
\textbf{\textit{User-side evaluation}} judges the quality of generated content based on users' satisfaction. The satisfaction can be collected either by explicit feedback or implicit feedback like in the traditional retrieval-based recommender paradigm. In detail, 1) explicit feedback includes users' ratings and conversational feedback, \eg ``I like this item'' in natural language. Moreover, we can design multiple facets to help users' evaluation, for instance, the style, length, and thumbnail for the evaluation of generated micro-videos. And 2) implicit feedback (\eg clicks) can also be evaluated. The widely used metrics such as the click-through rate, dwell time, and user retention rate are still applicable to measure users' satisfaction with the generated content.

\begin{figure}[t]
\setlength{\abovecaptionskip}{0.1cm}
\setlength{\belowcaptionskip}{-0.20cm}
\centering
\includegraphics[scale=0.72]{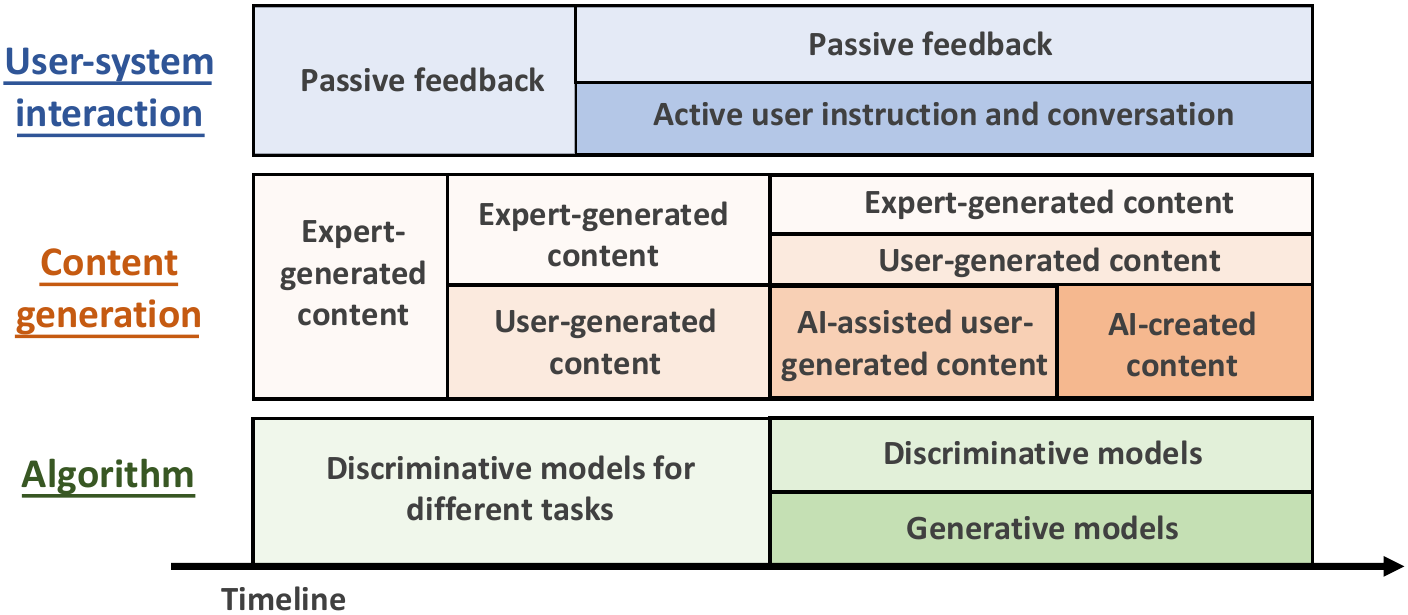}
\caption{A potential roadmap for the GeneRec paradigm from three aspects: user-system interaction, content generation, and recommender algorithm.}
\label{fig:roadmap}
\end{figure}

\vspace{3pt}
\noindent$\bullet$ \textbf{Roadmap.}
{In Figure~\ref{fig:roadmap}, we outline the future development roadmap of GeneRec by examining three separate aspects: user-system interaction, content generation, and recommender algorithms.} 
\begin{itemize}[leftmargin=*]
    \item [1)] {\textbf{User-system interaction.} 
    Users' passive feedback (\eg clicks) actually contains many subtle preferences, which might even be difficult for users themselves to articulate. Historically, recommender systems have heavily relied on user features and such passive feedback to model user preference. Beyond passive feedback, we believe that future recommender systems will be able to engage in multimodal conversations with users through voice, visual, and textual interactions, much like the AI assistant Jarvis\footnote{\url{https://en.wikipedia.org/wiki/J.A.R.V.I.S.}.}, to understand users' needs. By combining conversation with historical passive feedback, recommender systems can build more comprehensive user profiles, and subsequently, they can retrieve or generate content as recommendations to fulfill users' information needs. }

    {Existing LLMs have already achieved proficient textual conversation and burgeoning capabilities in multimodal interactions{~\cite{driess2023palm}}, effectively meeting the requirements for barrier-free communication with users. In the next years, a direction worth exploring is how to enable LLMs to quickly acquire and comprehend the item content in the recommendation domain. The new item content is continually emerging and the item popularity is quickly shifting over time. By efficiently understanding the item content, LLMs can better communicate with users and understand their information needs. Additionally, another direction to investigate in the coming years is how to combine active user conversation with passive feedback for more comprehensive user modeling.} 
    
    \item [2)] {\textbf{Content generation.} Content generation has three evolution phases. 1) Expert-generated professional content, such as carefully crafted movies and music. This type of content typically has high quality yet also comes with high production costs, making it challenging to generate large quantities quickly. 2) User-generated content, driven by the rise of micro-videos, has turned many users on the platform into content creators, greatly enriching content diversity and reach. However, user-generated content often varies widely in quality, with many items being of lower quality. 3) In the era of AIGC, AI can assist users in content creation and may even have the potential to independently generate new content. We can leverage AI to assist content creators and users themselves in content generation tasks, such as generating music for micro-videos or refining news articles. This helps improve the quality of user-generated content. During the phase of AI-assisted generation, users may still need to assume a significant portion of responsibility, including providing instructions and content design. If generative AI continues to advance, it is promising that generative AI can learn user preference directly from a vast amount of high-quality content and continuously create new content. In this way, AI-generated content will complement user-generated or expert-generated content to meet users' information needs. 
    }
    \item [3)] {\textbf{Algorithm.} Discriminative algorithms have traditionally been the mainstream approach for item rankings in recommendations, including but not limited to methods such as matrix factorization{~\cite{rendle2009bpr,wu2021self}}, graph neural networks-based methods{~\cite{he2020lightgcn,chang2021sequential}}, and transformer-based sequential algorithms{~\cite{kang2018self,xie2021adversarial,zhou2020s3}}. Generative models such as variational autoencoders have some explorations{~\cite{wang2022causal,liang2018variational}} while they receive relatively limited scrutiny. However, the recent emergence of LLMs has sparked a trend in using generative models for recommendation. Researchers are exploring the integration of various recommendation tasks, such as sequential recommendation, click-through rate prediction, recommendation explanations, and conversational recommendations, within a generative framework based on LLMs{~\cite{geng2022recommendation}}.
    Furthermore, they are also delving into how to better leverage word knowledge within LLMs for recommendation purposes{~\cite{lin2023multi}} and how to enable LLMs to effectively capture collaborative filtering signals in user-item interactions{~\cite{zhang2023collm}}. We believe that in the future, there will be a coexistence and mutual promotion of discriminative and generative models to accomplish various recommendation tasks. It is even possible that a unified LLM-based recommender model may emerge for various recommendation tasks, covering item retrieval, repurposing, and creation. 
    }
    
\end{itemize}

\section{Demonstration}
\label{sec:method}

\begin{figure}[t]
\setlength{\abovecaptionskip}{0.1cm}
\setlength{\belowcaptionskip}{-0.2cm}
\centering
\includegraphics[scale=0.5]{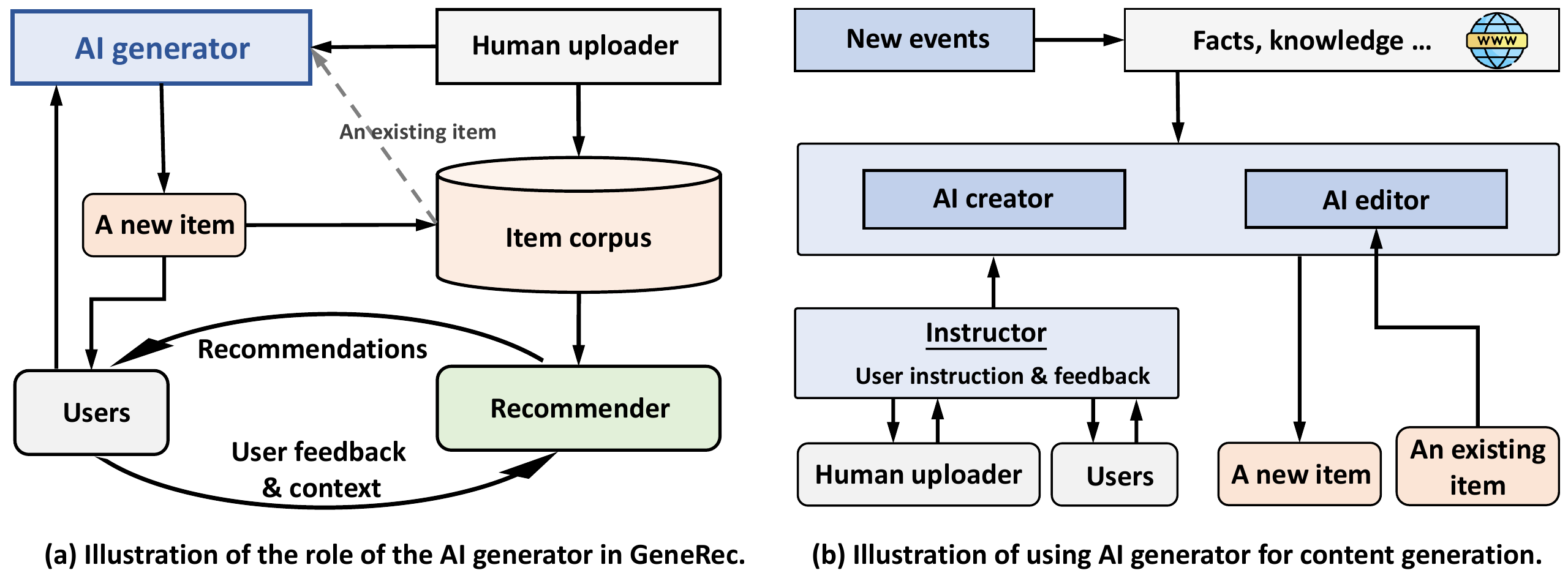}
\caption{A demonstration of GeneRec. Both users and human uploaders can interact with AI generators via instructions and feedback for content generation. 
The AI editor aims to repurpose existing items in the item corpus while the AI creator directly creates new items. The new item can be directly recommended to users or fed into the item corpus for item rankings.}
\label{fig:paradigm_demo}
\end{figure}

To instantiate the proposed GeneRec, we develop three modules: an instructor, an AI editor, and an AI creator. As described in Figure~\ref{fig:paradigm_demo}, the instructor is responsible for {initializing the content generation and} pre-processing user instructions, while the AI editor and AI creator implement the AI generator for personalized item editing and creation, respectively.
{Lastly, we present several application scenarios and potential research tasks.}

\vspace{-3pt}
\subsection{Instructor}
%
{The instructor aims to pre-process user instructions and feedback to initialize the AI generator and guide the content generation process.} 

\vspace{2pt}
\noindent\textbf{$\bullet$ Input:} Users' multimodal conversational instructions and the feedback over historically recommended items. 

\vspace{2pt}
\noindent\textbf{$\bullet$ Processing:} 
Given the inputs, the instructor may still need to engage in multi-turn interactions with the users to fully understand users' information needs. 
Thereafter, the instructor analyzes the multimodal instructions and user feedback to determine whether 
there is a need to initiate the AI generator to meet users' information needs. 
If the users have explicitly requested AIGC via instructions or rejected human-generated items many times, the instructor may enable the AI generator for content generation. 
The instructor then pre-processes users' instructions and feedback as guidance signals, according to the input requirements of the AI generator. For instance, some pre-trained language models may need appropriately designed prompts and diffusion models might require the extraction of guidance embeddings from users' instructions and historically liked item features.

\vspace{3pt}
\noindent\textbf{$\bullet$ Output:} 1) The decision on whether to initiate the AI generator, and 2) the guidance signals for content generation. 


\vspace{-3pt}
\subsection{AI Editor and AI Creator}
To implement the content generation, we formulate two modules: an AI editor and an AI creator. 

\vspace{-2mm}
\subsubsection{\textbf{AI editor for personalized item editing}}\label{sec:demo_AI_editor}

As depicted in Figure~\ref{fig:paradigm_demo}(b), the AI editor intends to refine and repurpose existing items (generated by either humans or AI) in the item corpus {according to the instructions and historical feedback from either human uploaders or users. Human uploaders such as video uploaders on YouTube can interact with the AI editor to generate different versions of items, which can be fed into the general item corpus for rankings. Users can also generate personalized items for themselves with the assistance of the AI editor.} 

\vspace{3pt}
\noindent\textbf{$\bullet$ Input:} 1) The guidance signals extracted from user instructions and feedback by the instructor, 2) an existing item in the corpus, and 3) the facts and knowledge from the Web data. 

\vspace{3pt}
\noindent\textbf{$\bullet$ Processing:} 
Given the input data, 
the AI editor leverages powerful neural networks to learn the users' information needs and preference, and then repurpose the input item accordingly. The ``facts and knowledge'' here can provide some factual events, generation skills, common knowledge, laws, and regulations to help generate accurate, safe, and legal items. {Note that the ``facts and knowledge'' can be updated with new events and regulations continually. 
For instance, based on the daily events, the AI editor can help to revise and generate personalized news reports for users. 
}


\vspace{3pt}
\noindent\textbf{$\bullet$ Output:} An edited item that better fulfills users' information preference than the original one.


\subsubsection{\textbf{AI creator for personalized item creation}}
Besides the AI editor, we also develop an AI creator to generate new items based on personalized instructions and feedback. 

\vspace{3pt}
\noindent\textbf{$\bullet$ Input:} 1) The guidance signals extracted from user instructions and feedback by the instructor, and 2) the facts and knowledge from the Web data. 

\vspace{3pt}
\noindent\textbf{$\bullet$ Processing:} Given the guidance signals, facts, and knowledge, the AI creator learns the users' information needs, and creates new items to fulfill users' needs. 
{As illustrated in Figure~\ref{fig:AIGC_example}, the AI creator may know users' preference for Jay Chou's songs from the user's historical feedback, determine the singer to Zhen Tian based on user instructions, and learn the singing skills of Zhen Tian on the Web to make a music video of ``Worldly Tavern” performed by Zhen Tian.} 

\vspace{3pt}
\noindent\textbf{$\bullet$ Output:} A new item that fulfills users' information needs. 

\subsection{{Domain Applications of GeneRec}}\label{sec:domain_cases}
{High-quality AIGC via GeneRec is applicable in multiple domains. The related domains include but are not limited to the recommendations of textual content (\eg news, articles, books, and financial reports), advertisements, videos (\eg micro-videos, animation, and movies), images (\eg art portraits and landscape), and audio (\eg music). 
The users and human producers can leverage the AI generators to produce extensive items such as movies, music, advertisements, and micro-videos. 
Furthermore, it is also viable to harness AI generators for the design of various products, including fashion apparel and accessories, with the potential for a seamless transition into the manufacturing process.}

{In detail, we present several examples in different domains.}

\begin{itemize}[leftmargin=*]
    \item {\textbf{News recommendation.} 
    Human news editors or users on the news platform can define rules by instructions for GeneRec to generate news. AI generators can also infer potential user preference by analyzing the feedback from editors' or users' historical interactions. Consequently, GeneRec can produce personalized news articles daily, taking into account newly occurring events. 
    After news generation, fidelity checks are needed to detect misleading headlines, misinformation, satire, biased and deceptive news~\cite{zhou2020survey}. }

    \item {\textbf{Fashion recommendation.} 
    GeneRec can assist fashion designers in creating various versions of fashion products or directly generate personalized items for users. The passive feedback of fashion designers and users, such as clicks, plays a crucial role by providing AI generators with valuable insights into their implicit clothing preferences. 
    AI-generated digital products with high value (\eg popular products liked by lots of people) can be sent to factories for customized production. 
    Currently, some fashion brands are exploring this direction, such as Tribute Brand\footnote{\url{https://www.tribute-brand.com/}.} and The Dematerialised\footnote{\url{https://thedematerialised.com/}.}. 
    In the fashion domain, fidelity checks should primarily focus on the realism of product details, ensuring image clarity, color accuracy, and high detail level while considering style consistency and compatibility.
    }
    \item {\textbf{Music recommendation.} 
    AI generators have demonstrated their ability to produce high-quality music{~\cite{yu2023musicagent}}. Under the GeneRec framework, to enhance the personalization of music generation, we should empower AI generators to learn users' implicit preferences for artists, lyrics, melodies, and singing styles from users' feedback. 
    In terms of fidelity checks, music generation should address copyright and legality concerns, ensuring compliance with all relevant regulations. 
    }
    \item {\textbf{Micro-video recommendation.} 
    Micro-video generation within the GeneRec framework is significantly challenging, as it involves the generation of multiple modalities, including textual subtitles, cover images, videos, background music, or ambient sounds. Nevertheless, AI-assisted user generation is a promising starting point. Extensive tools for AI-assisted content creation are emerging, encompassing features such as automatic music generation, cover image refinement, and subtitle generation. The field of AI-generated micro-videos is likely to transition progressively from AI-human collaboration toward AI-driven creation.
    Regarding fidelity checks, micro-video generation necessitates careful consideration of many aspects, including but not limited to bias, privacy, authenticity, safety, and realism. 
    }
\end{itemize}

{\subsection{Potential Research Tasks}}\label{sec:tasks}
Under the novel paradigm of GeneRec, there are various promising research tasks that deserve future exploration. We introduce some potential tasks as follows. 
\vspace{2pt}
\begin{itemize}[leftmargin=*]
    \item [1)] {
     \textbf{Instruction tuning for LLM-based instructor.} 
     One crucial factor in implementing the instructor of GeneRec is to enhance the instruction-following capabilities, allowing the instructor to comprehend user intention and take the right actions. These actions may involve generating responses, activating AI generators for content creation, or providing guidance for content generation. Existing work still cannot effectively achieve such an instructor. On one hand, conversational recommender models previously lack strong instruction-following capabilities, struggling to understand user intention. On the other hand, LLMs exhibit limited instruction-following abilities for item recommendations{~\cite{zhang2023recommendation,bao2023tallrec}}. 
     This limitation arises from a lack of instruction tuning within the recommendation domain. 
     Consequently, improving LLM-based instructors for GeneRec is an essential research task in the future. 
    }
    \item [2)] 
    {
    \textbf{Controlling for AI generator activation.}
    Under GeneRec, an essential task is to control whether to activate AI generators. Besides, after generating a new item, it needs to decide between directly recommending the generated item to users and ranking it with existing items. 
    These two decisions depend on the recommendation context and user behaviors, including users' instructions and feedback. For instance, if a user explicitly expresses a need for a generated item through instructions or consistently provides negative feedback on existing items, the AI generator should be initiated for content generation. 
    In the future, collecting diverse user behaviors and required actions to train the LLM-based instructor is vital for enhancing its decision-making capabilities. 
    }

    \item [3)]
    {
    \textbf{Personalized item editing.}
    One of the core tasks within GeneRec is the implementation of AI editors to assist human uploaders or users in personalized content editing. As mentioned in Section~\ref{sec:domain_cases}, item editing should be explored separately in specific domains. 
    While Generative AI has shown the ability to generate high-quality content in some domains, it is usually guided by user instructions. However, user instructions cannot describe various nuanced aspects of an item while crafting complex instructions is challenging and time-consuming. 
    Furthermore, many human uploaders and users may have implicit preferences that they themselves may not be aware of. As such, extracting user interests from noisy and implicit user feedback is helpful, which aligns with the central focus of recommendation models in recent years. 
    Hence, integrating users' implicit feedback into AI editors and effectively capturing users' implicit preference for personalized item editing holds great promise. 
    }
    \item [4)] 
    { 
    \textbf{Personalized item creation.}
    Going beyond personalized item editing, personalized item creation represents a more challenging task. This endeavor can begin with relatively easier domains, such as news and music recommendations, where generative AI technologies are relatively mature. 
    }
    \item [5)]
    {
    \textbf{Domain-specific fidelity checks.}
    As discussed in Section~\ref{sec:domain_cases}, implementing GeneRec in various domains requires careful consideration of domain-specific fidelity checks. It is of paramount importance to design fidelity evaluators tailored to specific domains. For instance, in news generation, special attention should be given to aspects like news authenticity and bias. 
    }
\end{itemize}

\section{Discussion}\label{sec:discussion}

{In this section, we present the comparison between the GeneRec paradigm and two related tasks:  conversational recommendation and traditional AI generation. Moreover, we present a possible vision for future developments of GeneRec. }

\subsection{Comparison}
{We illustrate how GeneRec differs from conversational recommendation and traditional AI generation tasks. }

\subsubsection{\textbf{Comparison with conversational recommendation}}
Conversational recommender systems rely on multi-turn natural language conversations with users to acquire user preference and provide recommendations~\cite{sun2018conversational,zhang2018towards,qu2018analyzing}. Although conversational recommender systems also consider multi-turn conversations, we highlight two critical differences with the GeneRec paradigm: 
1) {Previous conversational recommender models lack instruction-following abilities, leading to poor user experience.} The dramatic development of LLMs, especially ChatGPT, has brought a revolution to traditional conversational systems by significantly improving language understanding and generative abilities. {Based on LLMs, we can build a more powerful instructor for GeneRec.}
And 2) GeneRec automatically repurposes or creates items through the AI generator to meet users' specific information needs while conversational recommender systems are still retrieving human-generated items.

\subsubsection{\textbf{Comparison with traditional content generation}}
There exist various cross-modal AIGC tasks such as text, image, and audio generation conditional on users' images~\cite{yang2022pastiche}, single-turn queries~\cite{rombach2022high}, and multi-turn conversations~\cite{jiang2021talk}. 
Nevertheless, there are essential differences between traditional AIGC tasks and GeneRec. 1) {GeneRec can utilize the user modeling techniques from traditional recommender systems to capture implicit user preference. 
For example, GeneRec may leverage user feedback such as clicks and dwell time to dig out the implicit user preference that is not indicated through user instructions. Users may not be aware of their preference for a particular type of micro-videos, whereas their clicks and long dwell time on these micro-videos can indicate this preference. 
Learning implicit user preference from user features and behaviors, including long-term, short-term, noisy, and implicit user feedback, has been the core research in the recommendation domain in the past years. 
Many prior techniques in the recommendation domain can be used to exploit implicit user preference for GeneRec. In this case, GeneRec can consider both users' explicit instructions and implicit preference to complement each other. 
}
And 2) despite the success of AIGC, retrieving human-generated content remains indispensable in many cases to satisfy users' information needs. 
{
For example, if an emergency event occurs, journalists can send the latest video reports from the scene; besides, many human content producers have unique experiences or creativity that are difficult to replicate by generative AI. Therefore, compared to previous generative AI, GeneRec considers the cooperation between AI-generated and human-generated content to meet user information needs.
}
\subsection{A vision for future GeneRec.}
{In the future, generative AI might be incorporated into various information platforms to supplement human-generated content. 
It might start with using AI generators to assist human content producers in repurposing existing items or creating new items. With the technical advancement of AI generators, they may gradually begin to independently generate content in some simple scenarios. 
}
{Moreover, from a technical view, constrained by the development of existing AIGC technologies, we need to design different modules to achieve the generation tasks across multiple modalities.} However, we believe that building a unified AI model for multimodal content generation is a feasible and worthwhile endeavor{\cite{wu2023next,li2023multimodal}}. 
Under the GeneRec paradigm, we expect to see the increasing maturity of various generation tasks, along with growing integration between these tasks. 
Ultimately, with the inputs of user instructions and feedback, GeneRec will be able to perform retrieval, repurposing, and creation tasks by a unified model for personalized recommendations, leading to a totally new information seeking paradigm.


\section{Feasibility Study}
\label{sec:experiment}
To investigate the feasibility of instantiating GeneRec, we employ AIGC methods to implement some simple demos of the AI editor and AI creator {in micro-video application scenario} due to the widespread of micro-video content. 
{The instructor could be a ChatGPT-like interactive conversational user interface~\cite{gao2023chat}, an option-based interface for the user to choose ``like'' or ``dislike''~\cite{zhang2019proactive}, or a recommender model that collects the user interactions. 
The obtained user instructions or user feedback will then be preprocessed to guide the AI editor or AI creator to repurpose or create the micro-video content. In our experiments, we mainly focus on implementing the AI editor and AI creator so we simply use a recommender model MMGCN~\cite{wei2019mmgcn} as the instructor and obtain the user's historical interactions or the user embeddings from the well-trained MMGCN as the guidance.}


\vspace{3pt}
\noindent\textbf{$\bullet$ Dataset.} 
We utilize a high-quality micro-video dataset with raw videos.  
It contains $64,643$ interactions between $7,895$ users and $4,570$ micro-videos of diverse genres (\eg news and celebrities). 
The micro-video length is longer than eight seconds and each micro-video has a thumbnail with an approximate resolution of $1934\times1080$. 
We follow the micro-video pre-processing in~\cite{voleti2022MCVD} and each pre-processed micro-video has 400 frames with $64\times64$ resolution. 



\vspace{-5pt}
\subsection{\textbf{AI Editor}}
We design three tasks for personalized micro-video editing and separately tailor different methods for the tasks.

\vspace{-5pt}
\subsubsection{\textbf{Thumbnail selection and generation}} 
Considering that personalized thumbnails might better attract users to click on micro-videos~\cite{liu2015multi}, we devise the tasks of personalized thumbnail selection and generation to present a more attractive and personalized micro-video thumbnail for users. 

\begin{figure*}[t]
\setlength{\abovecaptionskip}{-0.10cm}
\setlength{\belowcaptionskip}{-0.40cm}
  \centering 
  \subfigure{
    \includegraphics[width=5.5in]{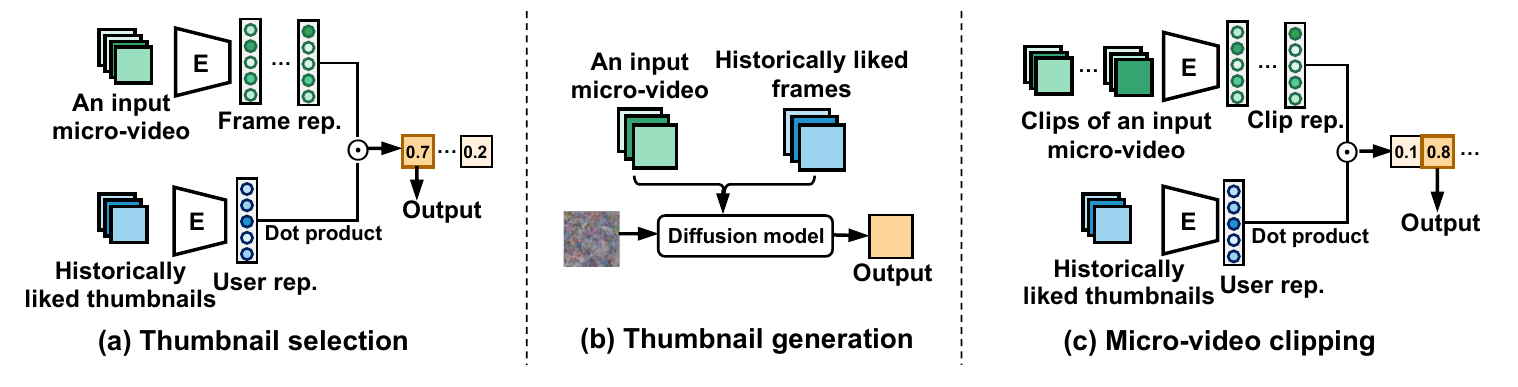}} 
\caption{Illustration of the implementation of editing tasks, including the process of AI editors for thumbnail selection, thumbnail generation, and micro-video clipping.}
  \label{fig:methods_fig}
\end{figure*}

\vspace{2pt}
\noindent\textbf{$\bullet$ Task.} We aim to generate personalized thumbnails based on user feedback without requiring user instructions. Formally, given a micro-video in the corpus and a user's historical feedback, the AI editor should select a frame from the micro-video as a thumbnail or generate a thumbnail to match the user's preference. 



\vspace{2pt}
\noindent\textbf{$\bullet$ Implementation.} To implement personalized thumbnail selection, we utilize the image encoder $f_\theta(\cdot)$ of a representative Contrastive Language Image Pre-training model (CLIP)~\cite{radford2021learning} to conduct zero-shot selection. 
As illustrated in Figure~\ref{fig:methods_fig}(a), given a set of $N$ frames $\{\bm{v}_i\}_{i=1}^{N}$ from a micro-video, and the set of $M$ thumbnails $\{\bm{t}_i\}_{i=1}^{M}$ from a user's historically liked micro-videos, we calculate 
\begin{equation}
\label{eq:thumbnail_selection}
\left \{
\begin{aligned}
& {\bm{t}^*} = \frac{1}{M}\sum_{i=1}^{M} f_{\theta}(\bm{t}_i), \\
& j=\mathop{\arg\max}\limits_{i\in\{1,\dots,N\}}\{{\bm{t}^{*T}}\cdot f_{\theta}(\bm{v}_i)\}, 
\end{aligned}
\right .
\end{equation}
where {${\bm{t}^*}$} is the average representation of $\{f_{\theta}(\bm{t}_i)\}_{i=1}^{M}$, and we select the $j$-th frame as the recommended thumbnail due to the highest dot product score between the user representation ${\bm{t}^*}$ and the $j$-th frame representation $f_{\theta}(\bm{v}_j)$. For performance comparison, we randomly select a frame from the micro-video (``Random Frame'') and utilize the original thumbnail (``Original'') as the two baselines.


To achieve personalized thumbnail generation, we adopt a newly pre-trained Retrieval-augmented Diffusion Model (RDM)~\cite{blattmann2022semi}, in which an external item corpus can be plugged as conditions to guide image synthesis. 
To generate personalized thumbnails for a micro-video, we combine this micro-video and the user's historically liked micro-videos as the input conditions of RDM (see Figure~\ref{fig:methods_fig}(b)).

\vspace{3pt}
\noindent\textbf{$\bullet$ Evaluation.} 
To evaluate the selected and generated thumbnails, we propose two metrics: 
1) \textbf{Cosine@$K$} that takes the average cosine similarity between the selected/generated thumbnails of $K$ recommended items and the user's historically liked thumbnails; and 2) \textbf{PS@$K$} that calculates the Prediction Score (PS) from a well trained MMGCN~\cite{wei2019mmgcn} by using the features of $K$ selected/generated thumbnails. In detail, we train an MMGCN by using the thumbnail features and user representations, and then averaging the prediction scores between the $K$ selected/generated thumbnails and the target user representation. 
Higher scores of Cosine@$K$ and PS@$K$ imply better results. For each user, we randomly choose $K=5$ or $10$ non-interacted items as recommendations and report the average results of ten experimental runs to ensure reliability. 

\begin{table}[t]
\setlength{\abovecaptionskip}{0.05cm}
\setlength{\belowcaptionskip}{0cm}
\caption{Performance of CLIP (thumbnail selection), RDM (thumbnail generation), and the baselines. The best results are highlighted in bold and the second-best underlined.} 

\label{tab:cover_selection}
\setlength{\tabcolsep}{2mm}{
\resizebox{0.56\textwidth}{!}{
\begin{tabular}{lcccc}
\hline
\multicolumn{5}{c}{\textbf{Thumbnail Selection and Generation}} \\ 
\multicolumn{1}{l}{} & \textbf{Cosine@5} & \textbf{Cosine@10} & \textbf{PS@5} & \textbf{PS@10} \\ \hline
\multicolumn{1}{l}{\textbf{Random Frame}} & 0.4796 & 0.4786 & 22.6735 & 23.1950 \\
\multicolumn{1}{l}{\textbf{Original}} & 0.4978 & 0.4970 & 22.2606 & 22.7445 \\ 
\multicolumn{1}{l}{\textbf{CLIP}} & {\ul 0.5142} & {\ul 0.5134} & {\ul 22.7682} & {\ul 23.2854} \\
\textbf{RDM} & \textbf{0.5369} & \textbf{0.5347} & \textbf{23.0145} & \textbf{23.3712} \\ \hline
\end{tabular}
}}
\end{table}


\vspace{3pt}
\noindent\textbf{$\bullet$ Results.} The results of thumbnail selection and generation \wrt Cosine@$K$ and PS@$K$ are reported in Table~\ref{tab:cover_selection}, from which we have the following findings.  
1) ``Original'' usually yields better Cosine@$K$ scores than ``Random Frame'' 
since the thumbnails manually selected by the uploaders are more appealing to users than random frames. 
2) CLIP outperforms ``Random Frame'' and ``Original'' by considering user feedback, validating the efficacy of personalized thumbnail selection. 
3) RDM achieves the best results, justifying the superiority of using diffusion models to generate personalized thumbnails. 
The superior results are reasonable since RDM can generate thumbnails beyond existing frames, leading to a better alignment with user preference. 


\begin{figure}[t]
\setlength{\abovecaptionskip}{0cm}
\setlength{\belowcaptionskip}{-0.10cm}
\centering 
\subfigure{
    \includegraphics[width=3.8in]{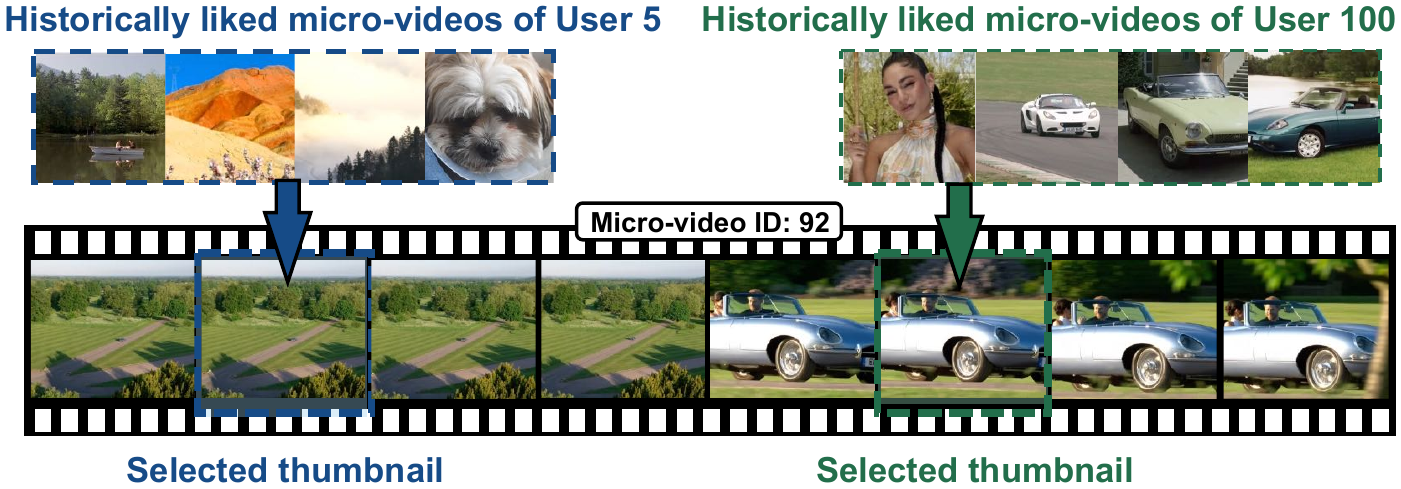}} 

\caption{Cases of personalized thumbnail selection by CLIP.}
  \label{fig:case_cover_selection}
\end{figure}

\begin{figure}[t]
\vspace{-0.3cm}
\setlength{\abovecaptionskip}{-0.10cm}
\setlength{\belowcaptionskip}{-0.50cm}
  \centering 
  \subfigure{
    \includegraphics[width=3.8in]{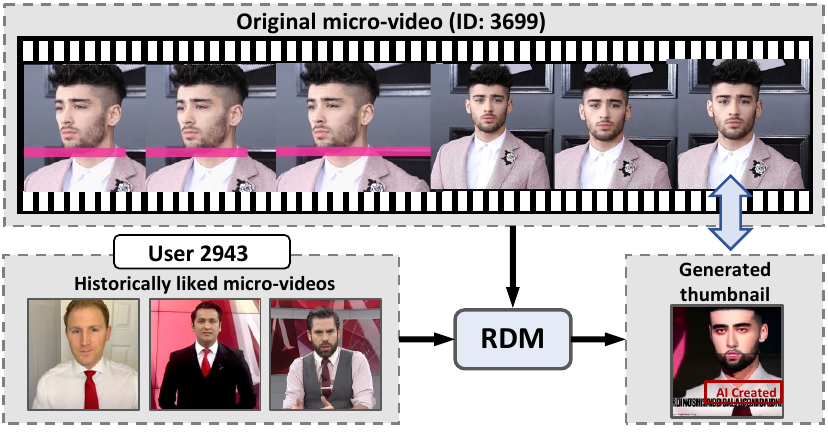}} 
\caption{Cases of personalized thumbnail generation.}
  \label{fig:case_cover_generation}
\end{figure}

\vspace{3pt}
\noindent\textbf{$\bullet$ Case study.} 
For intuitive understanding, we visualize several cases from CLIP and RDM. 
From the cases of CLIP in Figure~\ref{fig:case_cover_selection}, we observe that the second frame containing rural landscape is selected for User 5 due to the user's historical preference for magnificent natural scenery.
In contrast, the frame containing a fancy sports car is chosen for User 100 because this user likes to browse stylish vehicles. 
This reveals the effectiveness of CLIP in selecting personalized thumbnails according to different user preference.
The case of RDM for thumbnail generation is presented in Figure~\ref{fig:case_cover_generation}. From the generated result, we can find that RDM tries to insert some elements to the generated thumbnail to better align with user preference while maintaining key information of the original micro-video. 
For instance, RDM decorates the man with a white shirt and a red tie for User 2943 based on this user's historical preference. 
Such observation reveals the potential of using generative AI to repurpose existing items for meeting personalized user preference. 
Nevertheless, we can see that the generated thumbnail lacks fidelity to some extent, probably due to the domain gap between this micro-video dataset and the pre-training data of RDM. 

\subsubsection{\textbf{Micro-video clipping}} 
Given a long micro-video (\eg one longer than 1 minute), the task of personalized micro-video clipping aims to recommend only the users' preferred clip in order to save users' time and improve users' browsing experience~\cite{luo2022clip4clip}. 

\vspace{3pt}
\noindent\textbf{$\bullet$ Task.} Given an existing micro-video and a user's historical feedback, the AI editor needs to select a shorter clip comprising of a set of consecutive frames from the original micro-video as the personalized recommendation. 


\vspace{3pt}
\noindent\textbf{$\bullet$ Implementation.} 
Similar to the thumbnail selection, we leverage $f_\theta(\cdot)$ of CLIP~\cite{radford2021learning} to obtain personalized micro-video clips as shown in Figure~\ref{fig:methods_fig}(c). 
Given user representation ${\bm{t}^*}$ obtained from $M$ historically liked thumbnails via Eq. (\ref{eq:thumbnail_selection}), and a set of $C$ clips $\{\bm{c}_i\}_{i=1}^C$ where each clip $\bm{c}_i$ has $L$ consecutive frames\footnote{The frame number $L$ is a hyper-parameter for micro-video clipping. We tune it in $\{4,8,16,32\}$ and choose 8 due to its better scores \wrt Cosine@5.} $\{\bm{v}^i_a\}_{a=1}^L$, we compute
\vspace{-0.15cm}
\begin{equation}
\vspace{-0.15cm}
\left \{
\begin{aligned}
& {\bm{c}^*_i} = \frac{1}{L}\sum_{a=1}^{L} f_{\theta}(\bm{v}^i_a), \\
& j=\mathop{\arg\max}\limits_{i\in\{1,\dots,C\}}\{{\bm{t}^{*T}}\cdot {\bm{c}^*_i}\}, 
\end{aligned}
\right .
\end{equation}
where ${\bm{c}^*_i}$ denotes the clip representation calculated by averaging the $L$ frame representations, and we select $j$-th clip as the recommended one due to its highest similarity with the user representation ${\bm{t}^*}$. 
For performance comparison, we select a random clip (``Random''), the first clip with $1 \sim L$ frames (``1st Clip''), and the original unclipped micro-video (``Unclipped'') as baselines.


\begin{table}[t]
\setlength{\abovecaptionskip}{0cm}
\setlength{\belowcaptionskip}{0cm}
\caption{Performance comparison between the baselines and CLIP with personalized micro-video clipping.}
\label{tab:video_clipping}
\setlength{\tabcolsep}{3.5mm}{
\resizebox{0.56\textwidth}{!}{
\begin{tabular}{lcccc}
\hline
\multicolumn{5}{c}{\textbf{Micro-video Clipping}} \\
 & \textbf{Cosine@5} & \textbf{Cosine@10} & \textbf{PS@5} & \textbf{PS@10} \\ \hline
\textbf{Random} & 0.4864 & 0.4851 & 22.1483 & 23.1401 \\
\textbf{1st Clip} & 0.4910 & 0.4899 & 22.1509 & 23.1657 \\
\textbf{Unclipped} & 0.4969 & 0.4976 & 22.1685 & 23.1700 \\
\textbf{CLIP} & \textbf{0.5052} & \textbf{0.5038} & \textbf{22.1863} & \textbf{23.1758} \\ \hline
\end{tabular}
}}
\end{table}

\begin{figure}[t]
\setlength{\abovecaptionskip}{0cm}
\setlength{\belowcaptionskip}{0cm}
  \centering 
  \subfigure{
    \includegraphics[width=3.8in]{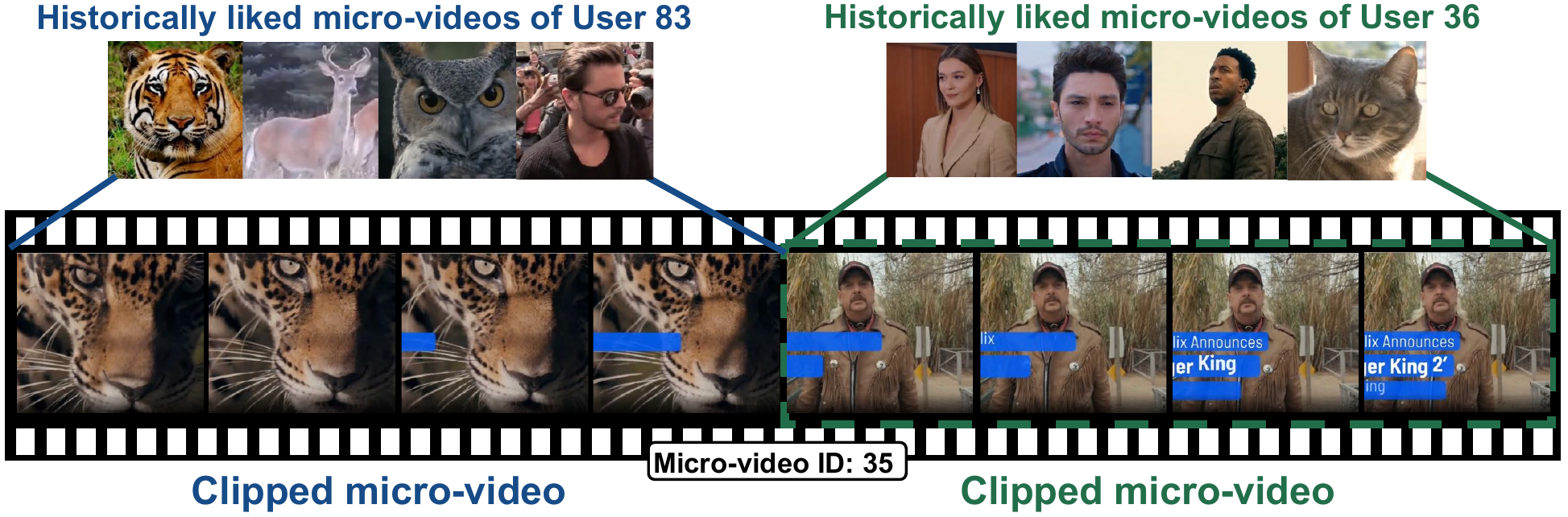}} 
\caption{Case study of micro-video clipping via CLIP~\cite{radford2021learning}.}
  \label{fig:case_video_clipping}
\end{figure}

\vspace{3pt}
\noindent\textbf{$\bullet$ Results.}
Similar to thumbnail selection and generation, we use Cosine@$K$ and PS@$K$ for evaluation by replacing thumbnails with frames to calculate cosine and prediction scores. 
From Table~\ref{tab:video_clipping}, we find that CLIP surpasses other approaches because it utilizes user feedback to select personalized clips that match users' specific preference. Besides, the superior performance of ``Unclipped'' over ``Random'' and ``1st Clip'' makes sense because the random clip and the first clip may lose some users' preferred content. 


\vspace{3pt}
\noindent\textbf{$\bullet$ Case study.} 
In Figure~\ref{fig:case_video_clipping}, CLIP chooses two clips from a raw micro-video for two users with different preference. For User 83, the clip with the tiger is selected because of this user's interest in wild animals; in contrast, the clip with a man facing the camera is chosen for User 36 due to this user's preference for portraits. 


\subsubsection{\textbf{Micro-video content editing}}\label{sec:exp_video_content_editing} 
Users might wish to repurpose and refine the micro-video content according to personalized user preference. 
As such, we implement the AI editor to edit the micro-video content for satisfying users' information needs. 

\vspace{3pt}
\noindent\textbf{$\bullet$ Task.} Given an existing micro-video in the corpus, user instructions, and user feedback\footnote{We do not consider the ``facts and knowledge'' in Section~\ref{sec:demo_AI_editor} to simplify the implementation, leaving the knowledge-enhanced implementation to future work.}, the AI editor is asked to repurpose and edit the micro-video content to meet user preference.



\begin{figure*}[t]
\setlength{\abovecaptionskip}{-0.10cm}
\setlength{\belowcaptionskip}{0cm}
  \centering 
  \subfigure{
    \includegraphics[width=4in]{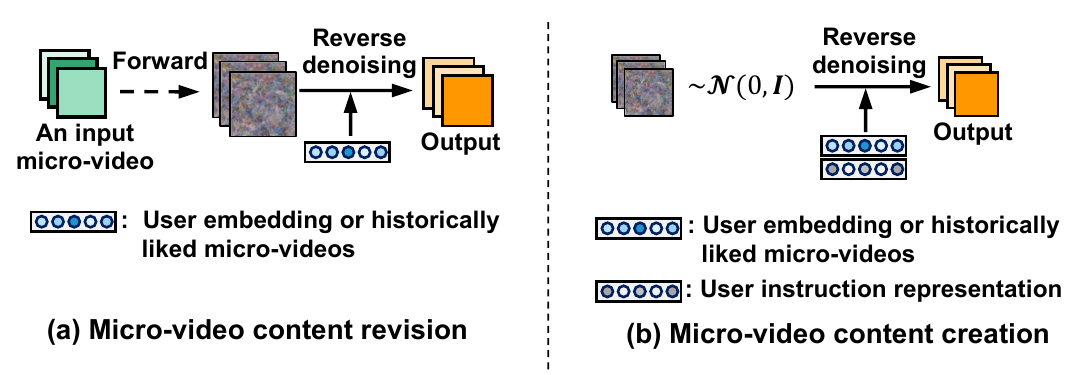}} 
\caption{Illustration of the implementation of editing and creation tasks. (a) depicts the procedure of AI editor for micro-video content revision and (b) shows the process of AI creator for micro-video content creation.}
  \label{fig:methods_fig2}
\end{figure*}

\vspace{3pt}
\noindent\textbf{$\bullet$ Implementation.} 
We consider two subtasks for micro-video content editing: 1) \textbf{\textit{micro-video style transfer}} based on user instructions, where we simulate user instructions to select some pre-defined styles and utilize an interactive tool VToonify\footnote{\url{https://github.com/williamyang1991/vtoonify/}.}~\cite{yang2022vtoonify} to achieve the style transfer; 
and 2) \textbf{\textit{micro-video content revision}} based on user feedback. We resort to a newly published Masked Conditional Video Diffusion model (MCVD)~\cite{voleti2022MCVD} for micro-video revision. 
The revision process is presented in Figure~\ref{fig:methods_fig2}(a). We first fine-tune MCVD on this micro-video dataset by reconstructing the users' liked micro-videos conditional on user feedback. 
During inference, we forwardly corrupt the input micro-video by gradually adding noises, and then perform step-by-step denoising to generate an edited micro-video guided by user feedback. 
The user feedback for fine-tuning and inference can be obtained from: 1) user embeddings from a pre-trained recommender model such as MMGCN (denoted as ``User\_Emb''), and 2) averaged features of the user's historically liked micro-videos (``User\_Hist''). 
To evaluate the quality of generated micro-videos, we follow~\cite{voleti2022MCVD} and adopt the widely used FVD metric~\cite{unterthiner2019fvd}, which measures the distribution gap between the real micro-videos and generated micro-videos. 
{Specifically, FVD builds on the principles underlying Frechet Inception Distance (FID~\cite{heusel2017gans}) and additionally considers the temporal coherence for videos. The FVD metric is formally written as:
\begin{equation}
    \text{FVD}=|\mu_R - \mu_G|^2 + \text{Tr}\left(\Sigma_R+\Sigma_G-2(\Sigma_R\Sigma_G)^{\frac{1}{2}}\right),
\end{equation}
where $\mu_R$ and $\mu_G$ are the means and $\Sigma_R$ and $\Sigma_G$ are the co-variance matrices of the distribution of feature representation of real-world videos and the generated videos. The feature representations are obtained by the pre-trained Inflated 3D Convnet (I3D~\cite{carreira2017quo}), which considers the temporal coherence of the visual content across a sequence of frames. 
}
A lower FVD score indicates higher quality. 
%
%

\begin{figure}[t]
\setlength{\abovecaptionskip}{0cm}
\setlength{\belowcaptionskip}{0cm}
  \centering 
  \subfigure{
    \includegraphics[width=2.8in]{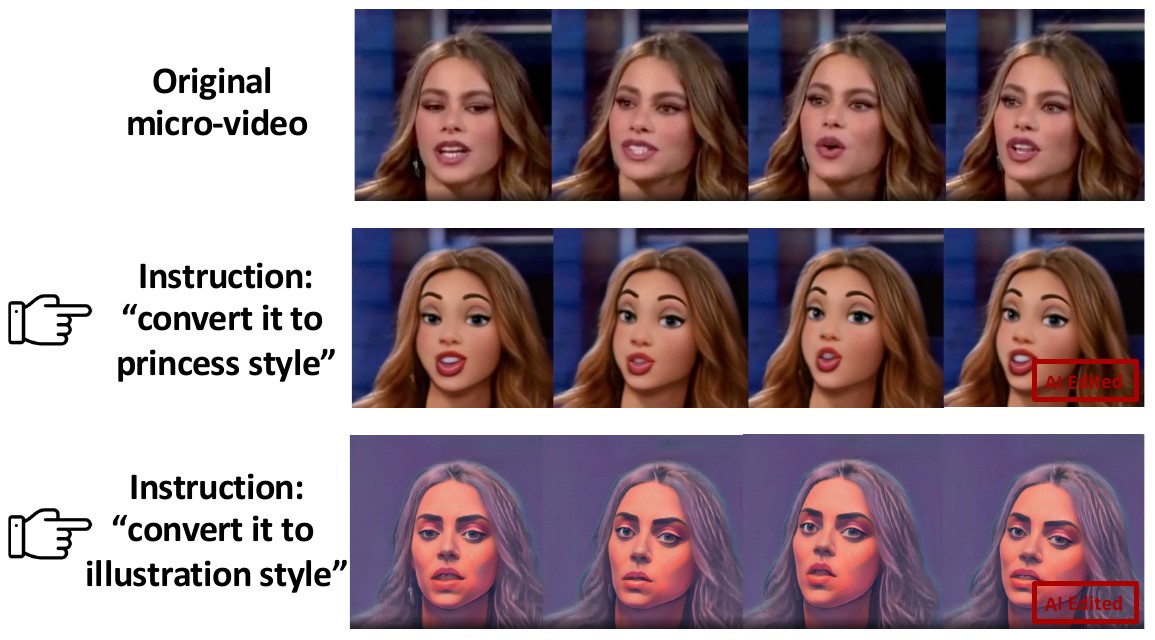}}   
\caption{Examples of personalized micro-video style transfer via VToonify.}
  \label{fig:case-video-gen-vtoonify}
\end{figure}


\begin{table}[t]
\setlength{\abovecaptionskip}{0cm}
\setlength{\belowcaptionskip}{0cm}
\caption{Quantitative results of MCVD based on User\_Hist and User\_Emb. As a reference, the FVD score of unconditional micro-video revision is 745.9443.} 
\label{tab:video_content_editing}
\setlength{\tabcolsep}{1mm}{
\resizebox{0.56\textwidth}{!}{
\begin{tabular}{lccccc}
\hline
\multicolumn{6}{c}{\textbf{Micro-video Content Revision}} \\
 & \textbf{Cosine@5} & \textbf{Cosine@10} & \textbf{PS@5} & \textbf{PS@10} & \textbf{FVD} \\ \hline
\textbf{Original} & 0.5010 & 0.5083 & 25.8900 & 24.6800 & - \\
\textbf{User\_Hist} & 0.5166 & 0.5127 & 25.9012 & 24.7107 & 783.7505 \\
\textbf{User\_Emb} & \textbf{0.5273} & \textbf{0.5200} & \textbf{26.0200} & \textbf{24.7900} & \textbf{646.7156} \\ \hline
\end{tabular}
}}
\end{table}

\begin{figure}[t]
\setlength{\abovecaptionskip}{0cm}
\setlength{\belowcaptionskip}{0cm}
  \centering 
  \subfigure{
    \includegraphics[width=3.8in]{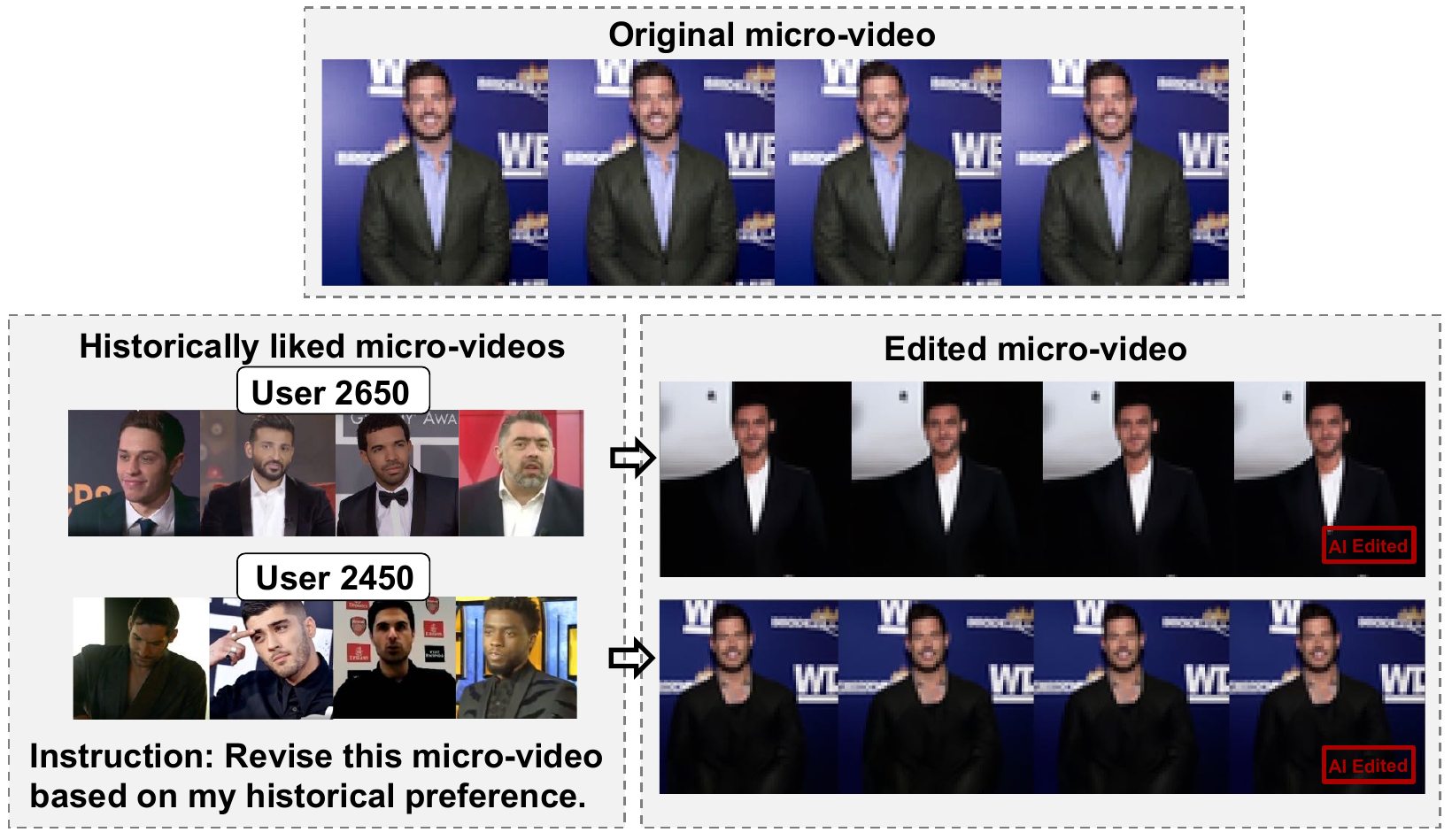}}   
\caption{Case study of personalized micro-video content revision via MCVD (User\_Emb).}
  \label{fig:case_video_content_editing}
\end{figure}

\vspace{3pt}
\noindent\textbf{$\bullet$ Results.} 
We show some cases of {micro-video style transfer} in Figure~\ref{fig:case-video-gen-vtoonify}. The same micro-video is transferred into different styles according to users' instructions. From Figure~\ref{fig:case-video-gen-vtoonify}, we can observe that the repurposed videos show high fidelity and quality, validating that the task of micro-video style transfer can be well accomplished by existing generative models. 
However, it might be costly for users to give detailed instructions for more complex tasks, thus considering user feedback as guidance, especially implicit feedback like in micro-video content revision, is worth studying.

Quantitative results of {micro-video content revision} are summarized in Table~\ref{tab:video_content_editing}, from which we have the following observation. 1) The edited items (``User\_Hist'' and ``User\_Emb'') cater more to user preference, validating the feasibility of generative models for personalized micro-video revision. 
2) ``User\_Emb'' outperforms ``User\_Hist'' possibly because the pre-trained user embeddings contain compressed critical information. 
``User\_Hist'' directly fuses the raw features from the user's historically liked micro-videos, which inevitably include some noises. 
And 3) the FVD score of ``User\_Emb'' is significantly smaller than the unconditional revision, indicating the high quality of the edited micro-videos.

In addition, we analyze the cases of two users in Figure~\ref{fig:case_video_content_editing}, where the original micro-video depicts a male standing in front of a backdrop. Given the users' instruction ``revise this micro-video based on my historical preference'', we initiate MCVD to repurpose the micro-video to meet personalized user preference. 
Specifically, since User 2650 prefers male portraits in a black suit and a white shirt, MCVD converts the dressing style of the man to match this user's preference. In contrast, for User 2450 who favors black shirts, MCVD alters both the suit and shirt to black accordingly. 
Despite the edited micro-videos having some deficiencies (\eg corrupted background for User 2650), we can find that integrating user instructions and feedback for higher-quality micro-video content revision is a promising direction. Besides, we highlight that the content revision and generation should add necessary watermarks and seek permission from all relevant stakeholders.

\subsection{\textbf{AI Creator}}
In this subsection, we explore instantiating the AI creator for micro-video content creation.

\vspace{-3pt}
\subsubsection{\textbf{Micro-video content creation}} Beyond repurposing thumbnails, clips, and content of existing micro-videos, we formulate an AI creator to create new micro-videos from scratch. 

\vspace{3pt}
\noindent\textbf{$\bullet$ Task.} Given the user instructions and the user feedback over historical recommendations, the AI creator aims to create new micro-videos to meet personalized user preference. 

\begin{figure}[t]
\setlength{\abovecaptionskip}{-0.1cm}
\setlength{\belowcaptionskip}{0cm}
  \centering 
  \subfigure{
    \includegraphics[width=3.2in]{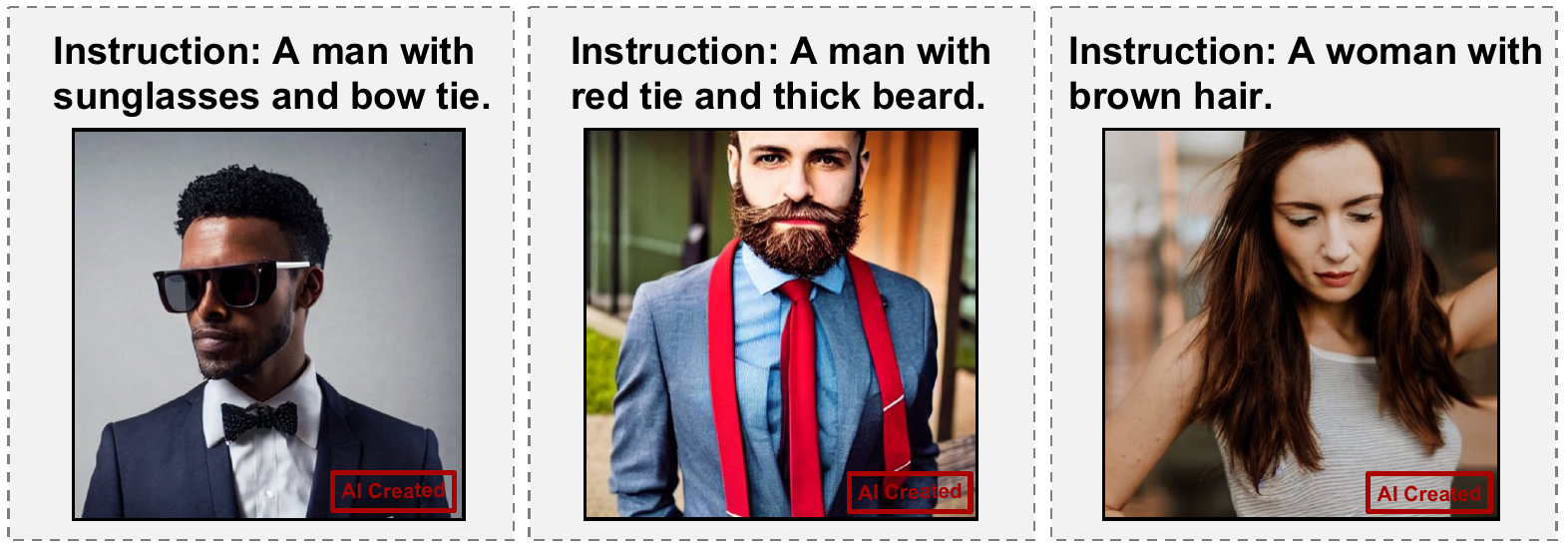}} 
\caption{Examples of single-turn instructions for text-to-image generation via stable diffusion.}
  \label{fig:text-to-image}
\end{figure}

\vspace{3pt}
\noindent\textbf{$\bullet$ Implementation.} 
Before implementing content creation, we investigate the performance of image synthesis based on user instructions, where we construct users' single-turn instructions and apply stable diffusion~\cite{rombach2022high} for image synthesis.
From the generated images in Figure~\ref{fig:text-to-image}, we can find that stable diffusion is capable of generating high-quality images according to users' single-turn instructions. 
Here, we explore the possibility of \textbf{\textit{micro-video content creation}} via the video diffusion model MCVD. 
As presented in Figure~\ref{fig:methods_fig2}(b), MCVD first samples a random noise from the standard normal distribution, and then it generates a micro-video based on personalized user instructions and user feedback through the denoising process. 
We write some textual single-turn instructions by humans such as ``a man with a beard is talking'', and encode them via CLIP~\cite{radford2021learning}. The encoded instruction representation is then combined with user feedback for the conditional generation of MCVD. 
To represent user feedback, we still employ the pre-trained user embeddings (``User\_Emb'') and the average features of the user's historically liked micro-videos (``User\_Hist''). Similar to micro-video content revision, we fine-tune MCVD on the micro-video dataset conditional on users' encoded instructions and feedback, and then apply it for content creation.

\vspace{3pt}
\noindent\textbf{$\bullet$ Results:} 
From Table~\ref{tab:video_content_generation}, we can find that generated micro-videos can achieve higher Cosine@$K$\footnote{We cannot calculate PS@K via MMGCN because the newly created items do not have item ID embeddings, which are necessary for MMGCN prediction.} values as the generation is guided by personalized user feedback and instructions. 
In spite of the high values of Cosine@K, the quality of generated micro-videos from ``User\_Emb'' and ``User\_Hist'' is worse than the unconditional creation as shown by the larger FVD scores. 
The case study in Figure~\ref{fig:case_video_content_generation}(a) also validates the unsatisfactory generation quality. Specifically, MCVD generates a micro-video containing a woman with long brown hair for User 11 where the women's face is distorted, {evident in the blurred rendering of the mouth and the generated hair extending to the chin region (see Figure~\ref{fig:case_video_content_generation}(b)).} 
And the generated portrait for User 38 is also blurred and distorted as shown in Figure~\ref{fig:case_video_content_generation}(a) and (b). {We can find that
the collar and coat are conjoined, and the left arm remains absent in the generated content.}


\begin{table}[t]
\setlength{\abovecaptionskip}{0cm}
\setlength{\belowcaptionskip}{-0cm}
\caption{Quantitative results of MCVD based on User\_Hist and User\_Emb. As a reference, the FVD score of unconditional micro-video creation is 727.8236.}
\label{tab:video_content_generation}
\setlength{\tabcolsep}{4mm}{
\resizebox{0.56\textwidth}{!}{
\begin{tabular}{lccc}
\hline
\multicolumn{4}{c}{\textbf{Micro-video Content Creation}} \\
 & \textbf{Cosine@5} & \textbf{Cosine@10} & \textbf{FVD} \\ \hline
\textbf{Original} & 0.4883 & 0.4907 & - \\
\textbf{User\_Hist} & 0.4902 & 0.4915 & \textbf{735.0413} \\
\textbf{User\_Emb} & \textbf{0.5356} & \textbf{0.5376} & 743.1090 \\ \hline
\end{tabular}
}}
\end{table}

\begin{figure}[t]
\setlength{\abovecaptionskip}{0cm}
\setlength{\belowcaptionskip}{0cm}
\centering 
\subfigure{
\includegraphics[width=4in]{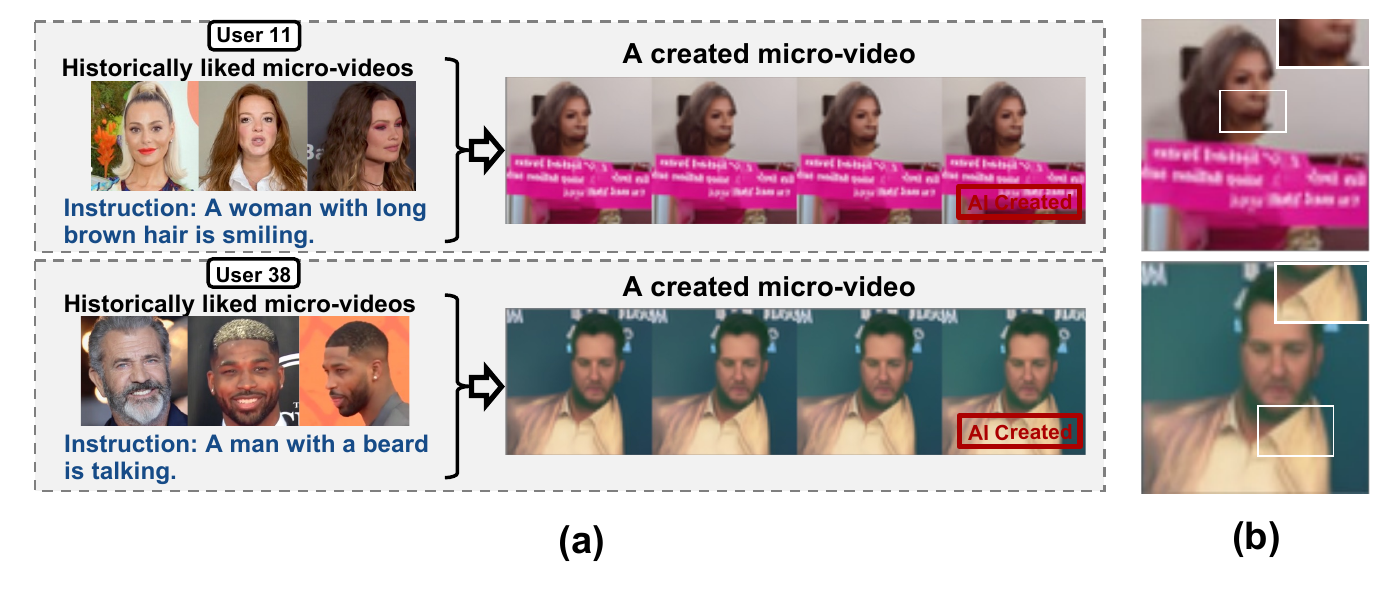}} 
\caption{Case study of personalized micro-video content creation via MCVD (User\_Emb).}
\label{fig:case_video_content_generation}
\vspace{-0.40cm}
\end{figure}


Admittedly, the current results of personalized micro-video creation are less satisfactory. This inferior performance might be attributed to that: 
1) the simple single-turn instructions fail to describe all the details in the micro-video; and 2) the current micro-video dataset lacks sufficient micro-videos, facts, and knowledge, thus limiting the creative ability of the AI generator. 
In this light, it is promising to enhance the AI generator in future work from {three} aspects: 
1) enabling the powerful ChatGPT-like tools\footnote{In this work, we do not explore the usage of ChatGPT because it was recently released and we have insufficient time for comprehensive exploration.} to implement the instructor and acquire detailed user instructions; 
{
2) pursuing more comprehensive user modeling and better encoding users' implicit preference to AI generators; and 
3) using more advanced algorithms of generative AI with strong prior world knowledge through pre-training on more data in different modalities, such as image (\eg Midjourney), video (\eg PikaLab), and audio (\eg AudioCraft). 
}
We believe that the development of video generation might also follow the trajectory of image generation (see Figure \ref{fig:text-to-image}) and eventually achieve satisfactory generation results.

{
Moreover, we emphasize the importance of addressing copyright issues regarding AI-generated content. For GeneRec in the micro-video domain, we offer potential solutions from two perspectives: 
1) authorship belongs to the platform when the micro-video is fully generated by the AI creator; 
2) the uploaders hold the authorship of the generated micro-video if the generated content is edited from the uploader-created content. 
Nevertheless, the policy for tackling copyright violations needs formal regulations, especially from the government's view. 
}

\vspace{3pt}
\noindent\textbf{$\bullet$ Running example of GeneRec on micro-video recommendation.} 
{
To run the whole GeneRec paradigm for micro-video recommendation, it involves the steps as follows: 
1) A user watches the recommended micro-videos, or directly searches or expresses what they prefer to watch (\ie information needs) at the moment; 
2) Instructor with conversational interaction interface acquires the user's information needs and pre-processes instructions. 
3) Based on the pre-processed user's instructions and historical feedback, the instructor will decide whether to use the AI editor or the AI creator for generating personalized micro-videos. For example, if the user requests the instructor to provide creative AI-generated content, the AI creator will then be called to generate a new micro-video as the response. 
4) Post-processing of the micro-video such as the checks of quality, relevance, and fidelity will then be conducted. 
5) Lastly, the instructor will decide to directly recommend the AI-generated micro-video to the user, or rank the AI-generated micro-video with all existing micro-videos in the item corpus. 
6) Provide the final recommendation and go back to step 1) or 2) based on the user's information needs.
}
\section{Related Work}
\label{sec:related_work}

\noindent\textbf{$\bullet$ Recommender Systems.}
The traditional retrieval-based recommender paradigm constructs a loop between the recommender models and users~\cite{ren2022variational,zou2022improving}. Recommender models rank the items in a corpus and recommend the top-ranked items to the users~\cite{zehlike2022fair,hu2022learning}. The collected user feedback and context are then used to optimize the next round of recommendations.
Following such paradigm, recommender systems have been widely investigated. Technically speaking, the most representative method is Collaborative Filtering (CF), which assumes that users with similar behaviors share similar preference~\cite{He2017Neural,ebesu2018collaborative,konstas2009social}. 
Early approaches directly utilize collaborative behaviors of similar users (\ie user-based CF) or items (\ie item-based CF). Later on, MF~\cite{rendle2009bpr} decomposes the interaction matrix into user and item matrices separately, laying the foundation for subsequent neural CF methods~\cite{He2017Neural} and graph-based CF methods~\cite{he2020lightgcn}. 
Beyond purely using user-item interactions, prior work considers incorporating context~\cite{rendle2011fast} and the content features of users and items~\cite{wei2019mmgcn,guy2010social} for session-based, sequential, and multimedia recommendations~\cite{Hidasi2016session, kang2018self,deng2021unified}. 
In recent years, various new recommender frameworks have been proposed, such as conversational recommender systems~\cite{sun2018conversational,tu2022conversational} acquiring user preference via conversations and user-controllable recommender systems~\cite{wang2022user} for controlling the attributes of recommended items. Recently, some researchers have considered using large language models for recommendations~\cite{bao2023tallrec,hou2023large,dai2023uncovering,li2023gpt4rec,liu2023chatgpt} while they mainly enhance recommender algorithms instead of producing content through generative AI or revolutionizing the user-system interaction interfaces. 



Past work only recommends human-generated items, which might fail to satisfy users' diverse information needs.  
In our work, we propose to empower traditional recommender paradigms with the ability of content generation to meet users' information needs and present a novel generative paradigm for next-generation recommender systems. 
{Satisfying personalized information needs has been studied in the Human-Computer Interaction (HCI) domain~\cite{arazy2015personalityzation}. While studies from HCI mainly focus on user modeling based on device- and context-specific information~\cite{volkel2019opportunities}, we emphasize the harness of ChatGPT-like models for advanced user-system interactions, supplementing traditional user feedback with better user engagement. }


\begin{figure}[t]
\setlength{\abovecaptionskip}{0cm}
\setlength{\belowcaptionskip}{-0.5cm}
\centering
\includegraphics[scale=0.66]{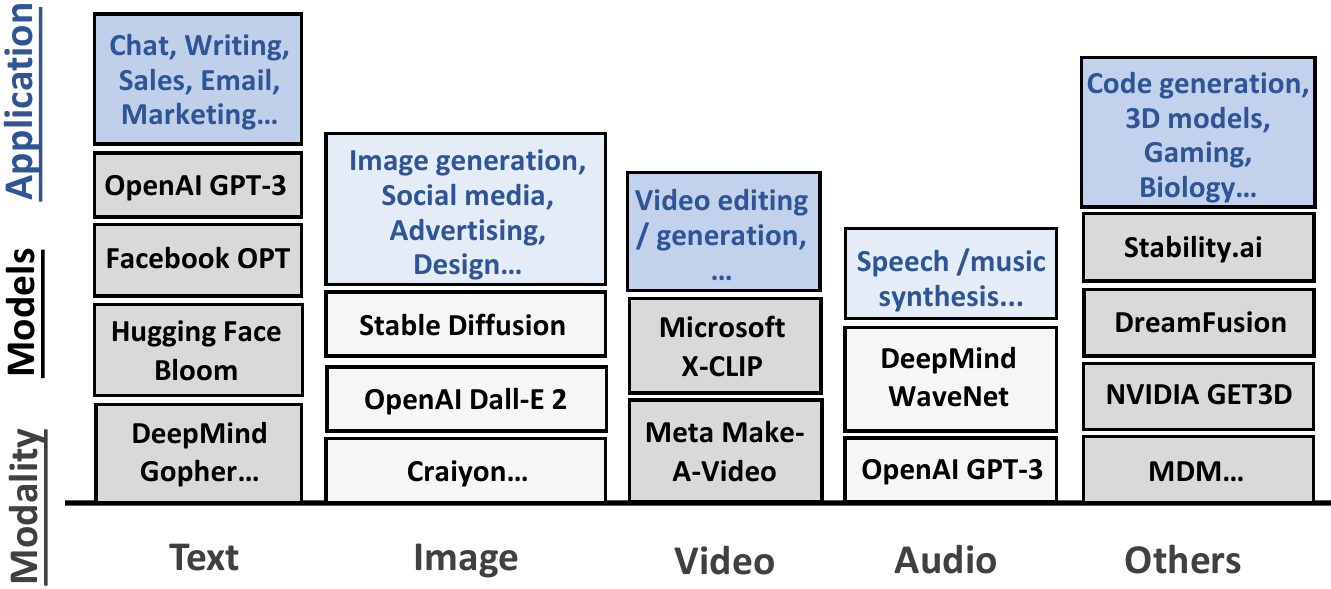}
\caption{Illustration of advanced models and potential applications of AIGC across different modalities.}
\label{fig:AIGC_landscape}
\end{figure}

\vspace{5pt}
\noindent\textbf{$\bullet$ Generative AI.}
The development of content generation roughly falls into three stages. 
At the early stage, most platforms heavily rely on high-quality professionally-generated content, which is however challenging to meet the demand of large-scale content production due to the high costs of professional experts. 
Later on, User-Generated Content (UGC) becomes prevalent due to the emergence of smartphones and well-wrapped generation tools. Despite its increasing growth, the quality of UGC is usually not guaranteed. 
Beyond human-generated content, recent years have witnessed third-stage content generation with groundbreaking generative AI techniques, leading to various AIGC-driven applications. 


As the core of AIGC, generative AI has been extensively investigated across a diverse range of applications as depicted in Figure~\ref{fig:AIGC_landscape}.
In the text synthesis domain, substantial methods are proposed to serve different tasks such as article writing and dialog systems. 
For example, the newly published ChatGPT demonstrates an impressive ability for conversational interactions. 
Image and video synthesis are also two prevailing tasks of AIGC. Recent advancement of diffusion models shows promising results in generating high-quality images in various aesthetic styles~\cite{dhariwal2021diffusion,nichol2021improved,song2020denoising} as well as high-coherence videos~\cite{ho2022video,voleti2022MCVD,hong2022cogvideo}. 
Besides, previous work has explored audio synthesis~\cite{ren2020popmag}. For instance, \cite{donahue2020end} proposes a generative model that aligns text with waves, endowing high-fidelity text-to-speech generation.
Furthermore, extensive generative models are crafted for other fields, such as 3D generation~\cite{guo2020action2motion}, gaming~\cite{de2020evolutionary}, and chemistry~\cite{polykovskiy2020molecular}. 

The revolution of generative AI has catalyzed the production of high-quality content and brought a promising future for generative recommender systems. The advent of AIGC can complement existing human-generated content to better satisfy users' information needs. 
Besides, the powerful generative language models can help acquire users' information needs via multimodal conversations. 


\section{Conclusion and Future Work}
\label{sec:conclusion}

In this work, we empowered recommender systems with the abilities of content generation and instruction guidance. In particular, we proposed a GeneRec paradigm, which could: 1) acquire users' information needs via user instructions and feedback, and 2) achieve both item retrieval, repurposing, and creation to meet users' information needs. To instantiate GeneRec, we formulated three modules: an instructor for pre-processing user instructions and feedback, an AI editor for repurposing existing items, and an AI creator for creating new items. Besides, we highlighted the importance of multiple fidelity checks to ensure the trustworthiness of the generated content, and {also pointed out the roadmap, application scenarios, and future research tasks of GeneRec.} We explored the feasibility of implementing GeneRec on micro-video generation and the experiments reveal some weaknesses and promising results of existing AIGC methods on various tasks.

This work formulates a new generative paradigm for next-generation recommender systems, leaving many valuable research directions for future work. In particular, 
1) it is critical to learn users' information needs from users' multimodal instructions and feedback. In detail, GeneRec should learn to ask questions for efficient information acquisition, reduce the modality gap to understand users' multimodal instructions, and integrate user feedback to complement instructions for better generation guidance. 
2) Developing more powerful generative modules for various tasks (\eg thumbnail generation and micro-video creation) is essential. Besides, we might implement some generation tasks through a unified model, where multiple tasks may promote each other. 
And 3) we should devise new metrics, standards, and technologies to enrich the evaluation and fidelity checks of AIGC. It is a promising direction to introduce human-machine collaboration for GeneRec evaluation and various fidelity checks. 






{
\tiny
\bibliographystyle{ACM-Reference-Format}
\balance
\bibliography{bibtex}
}


\end{document}